# Light-induced mass transport in amorphous chalcogenides: towards optical nanolithography and optical-field nanoimaging

**M.L.Trunov*,1,2, P.M. Lytvyn**3, P.M. Nagy4, A. Csik5, V.M. Rubish2, S. Kökényesi6**

[1] Uzhgorod National University, Narodna sq. 3, 88000, Uzhgorod, Ukraine
[2] Uzhgorod Scientific-Technological Center of IIR NAS Ukraine, Zamkovi shody st. 4a, 88000, Uzhgorod, Ukraine
[3] V. Lashkaryov Institute of Semiconductors NAS Ukraine, Nauky pr. 41, 03028, Kiev, Ukraine
[4] Research Centre for Natural Science, HAS, Pusztaszeri st. 59-67, H-1025, Budapest, Hungary
[5] Institute of Nuclear Research, HAS, Bem sq. 18/a, H-4026, Debrecen, Hungary
[6] University of Debrecen, Bem sq. 18/c, H-4026, Debrecen, Hungary

Two types of amorphous functional materials, based on light-sensitive inorganic compounds like Se and $As_{20}Se_{80}$ chalcogenide glass (ChG) were investigated with the aim to establish the influence of plasmonic fields, excited by the band-gap light in nanocomposite layers made of these compounds and gold nanoparticles on their photomechanical response. Both these basic materials are characterized by pronounced photoplastic effect and used for real-time optical recording of optoelectronic elements (based mainly on surface relief gratings) due to high photofluidity and polarization-dependent mass-transport. We established that mass-transport processes in these ChG can be enhanced in the presence of localized plasmonic fields generated by light if the condition of surface plasmon resonance (SPR) is fulfilled. The subjects of special interest are the mass-transport processes at the nano-scale stimulated in the nano-composite layers of either by the uniform or periodically distributed optical fields. It was found that irradiation by light with SPR really enhance the efficiency of mass-transport and produce surface nanostructurizations. The variation in the topography follows closely and permanently the underlying near field intensity pattern.

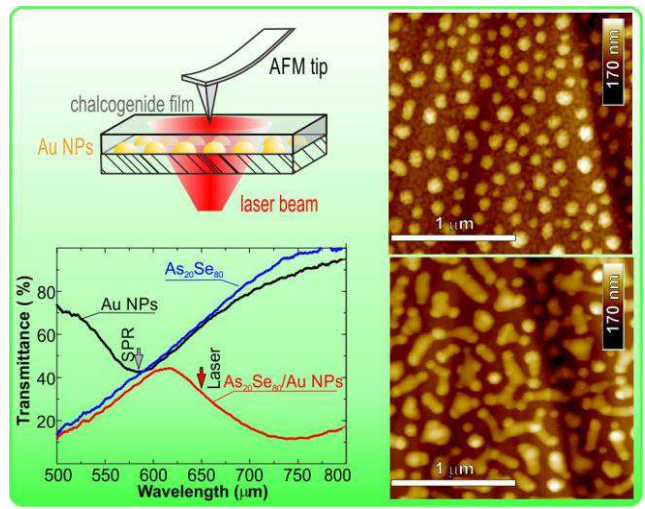

Nanostructurization of photosensitive amorphous $As_{20}Se_{80}$ film by surface plasmon near field irradiation.

**1 Introduction** Reversible local alteration of the chemical composition, optical properties or surface topography of thin (50-100nm) layer properties of thin films, are widely used in a modern nanoscale technique, e.g. in a sensor devices, opto-mechanical elements, etc. The control of these properties in a nanoscale requires the development of structures with "smart" surface that could be easily adjusted reversibly under external stimuli as local heating, e-beam or light illumination.

Such class of substances includes chalcogenide glassy semiconductors because their properties can easily be modified by irradiation with light from their absorption edge and can be restored following by thermal annealing. Besides the development of the methods for deposition of thin films on large area, it opens up new possibilities of their practical applications, such as optical storage drives with high-density recording, fast and durable photosensitive drums for photo-copying and photoresists for lithography, the elements of infrared laser equipment, holograms and elements of integrated optics, electronic switches, etc. Significant attention has been recently paid to the possibility of using chalcogenide glasses (ChG) in nanoplasmonics [1-3]. However, proposed devices [4], as a rule, are based on the Kretschmann scheme and use ordinary well-known photoinduced changes of optical parameters of ChG (the refractive index, absorption coefficient, etc.) omitting other options, in particular, photoinduced surface modification (patterning).

The prevailing opinion for a long time was that the modification of the surface of ChG films due to their irradiation with light was only possible with their additional treatment using chemical etching [5]. The discovery of the effect of photoinduced mass transport [6] under the influence of light at the absorption edge allowed obtaining surface topography purely by means of optical method and even at relatively low intensities of light waves [7-9]. This method typically uses a projection of the interference pattern that is formed by the interaction of two or more coherent plane waves on the surface of ChG film [10,11].

According to our earlier data, photoplastic effect in ChG films of certain compositions [12] accompanied by giant mass-transport of film material into a more or less lighted area (depending on the composition of ChG), which in its turn leads to the formation of a surface relief of a submicron amplitude [13]. The topographical changes of a surface which are obtained by using this method are limited by the lateral size due to the phenomenon of diffraction of light, which does not allow manufacturing of working elements of nanometer scale.

One of the ways to overcome the diffraction barrier can be the use of near light field (near-field) of metallic nanoparticles (NPs) integrated into a ChG film as plasmonic nanostructures [14]. Generation of localized plasmons in the noble NPs is widely used for enhancement of interaction of light with a matrix surrounding these plasmonic nanostructures. Incident light, absorbed by NPs and being converted into collective oscillations of free electrons in the NPs leads to a strong enhancement of the local electric field. This phenomenon titled as a surface plasmon resonance (SPR) occurs in case of noble NPs in visible spectral region and can be considered as generation of evanescent photons in the near-field region [15-16]. Consequently, the same nanoscale region of surrounding glass matrix should be excited through generation of carriers in case if the wavelength of SPR is within the spectral range of absorption edge of ChG. Such a carrier generation enhancement is widely used for improving the efficiency of the solar cells [17]. It is known also that electron-hole pair generation under band gap excitation is the general part of the mechanism of mass-transport process in ChG [18]. That is why it can be assumed that the local electric field enhancement can result in a significant performance improvement of mass-transport in a nanoscale.

At the same time such a structure (chalcogenide glass film / noble NPs) can also be used for mapping the distribution of the electromagnetic field of surface plasmons. It is also evident that the spatial distribution of the field can be controlled by appropriate changes in the size and geometry of NPs and/or by the change of their relative position. Thus, considerably more complex surface relief than by conventional interference lithography can be formed due to the interference of the light field of such plasmons and the change of the wavelength or polarization of irradiating light.

The main idea of this contribution is to investigate the influence of near-field irradiation on the formation of surface reliefs in ChG films with integrated Au NPs under the appropriate excitation of SPR by means of laser radiation. Instead of Kretschmann configuration that employs the variation of optical properties of the films, special case photoplastic effect (mass-transport) in amorphous chalcogenides will be used. Direct photoinduced fabrication of the reliefs with different shape and scale on film surface by lateral mass-transport under sub-wavelength surface plasmon near-fields is expected. It will allow us nanopatterning of the photosensitive films having the main advantage of plasmon nanolithography (to overcome the diffraction limit). On the other hand, more complex surface reliefs can be generated in a single step due to the possibility to tune the shape of the plasmon near-field radiation by adjusting the geometry of NPs. Another feature will be the mapping of surface plasmon intensity distribution.

**2 Materials and Experiment**
**2.1 Fabrication of NPs array** Using rapid thermal annealing, an array of randomly arranged Au NPs was formed on sapphire or Corning glass substrates with conductive indium tin oxide (ITO) layer. In the first stage, thin (10-20 nm) films of gold were condensed on the substrates by method of thermal evaporation. The subsequent annealing in argon atmosphere at temperatures of 400-700 °C led to the formation of randomly distributed hemispherical Au NPs in a diameter from 20 to 150 nm which reveal the SPR in the vicinity of 520-580 nm. The surface morphologies of the annealed thin films of gold with different initial thicknesses and thermal prehistory were characterized by AFM and shown in Fig. 1 (a)–(b). These morphologies clearly reveal the formation of NPs in annealed films. The optical transmission spectra of the prepared samples were measured with Ocean Optics spectrophotometer. It was shown that the SPR peak depends on the average size of Au NPs produced on the glass substrates (Fig. 1 c, curve 1-3).

**2.2 Photosensitive to SPR composite preparation** In the second stage, ChG films with the thickness of 30-300 nm were applied on top of the Au NPs with the use of thermal evaporation using a deposition rate of 1.5–3 nm·s$^{−1}$. The thickness of ChG films was measured with AFM on control samples (without Au NPs as sublayers).

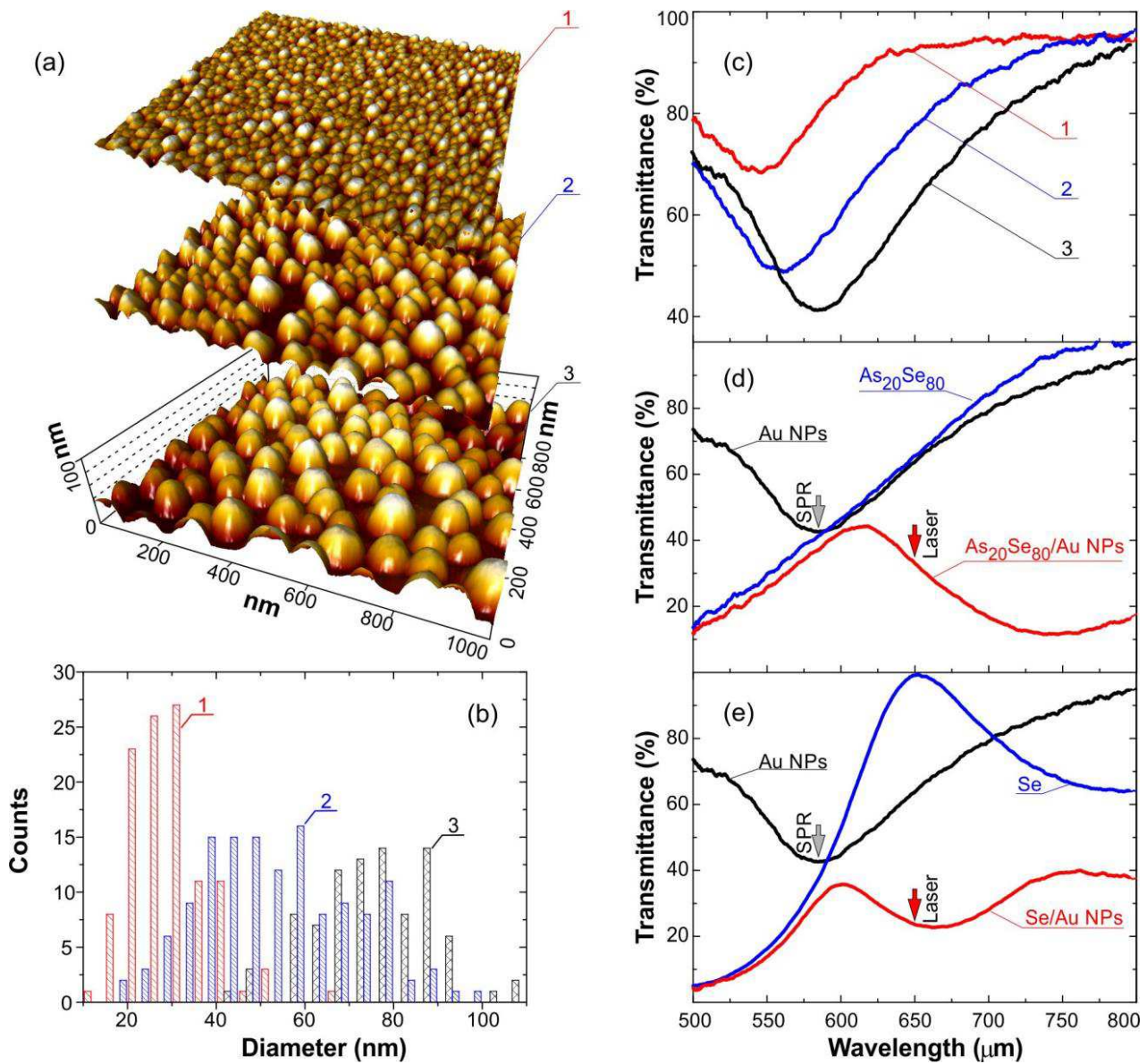

**Figure 1** AFM images of Au NPs with different diameters on the sapphire or glass substrates covered by ITO (a), corresponding size histograms (b) and optical transmittance spectra of appropriate Au NPs (c) and photosensitive structures: ChG film and ChG film /Au NPs composites (d-e). The compositions are indicated directly on the images. Thickness of the ChG film is 100 nm; the Au NPs corresponds to array No 3 in (a).Arrows show the initial position of SPR and the wavelength of the laser source that was used for irradiation of composite structures.

Note that as plasmon-induced effect on the ChG film structure is localized on the distances of near field existence in the range of 20-100 nm we should limit the thickness of ChG layer by 100 nm in Au NPs/ChG film composite. In some experiments devoted mostly to surface relief formation in the composite due to reason which will explain further we also used the ChG film with the thickness of 300 nm. The resulting structures of the samples and their optical transmittance (both as a pure film with a thickness of 100 nm and being deposited on Au NPs with average size of 40-80 nm) are shown in the Fig. 1 (d)-(e). As the film forming compositions were selected the following compounds of ChG: amorphous Se (a-Se) and $As_{20}Se_{80}$. The presence of significant mass-transport in $As_{20}Se_{80}$ films [19] served as the criterion for its choice; while a-Se was of interest due to the fact that it is the only ChG composition with the opposite direction of motion of the material (away from light) [20].

The Au NPs that were used for excitation of near-field illumination through localized surface plasmons satisfy the conditions for SPR excitation in the visible spectral region

(as shown by arrow in the Au NPs curves in Fig 1 (d)-(e)) where most of the basic ChG of As-S(Se) system has maximum to photoinduced response. However, if those NPs are covered by thin ChG layer the SPR spectra changed according to the differences in the refractive index, thickness of ChG layer and moreover, further shifted due to the photostimulated changes of its optical parameters. Such a way we obtained an efficient method of influencing the light induced structural transformations of the ChG film and measuring the kinetics of the process. It is obvious also that to achieve the objectives of the work, the frequency of the incident beam that causes the SPR must coincide with absorption band of ChG, or at least, be in its vicinity. That is why we used *in-situ* measurements of transformation of SPR spectra during amorphous layer deposition and interrupts the process after reaching the appropriate thickness of the film (see supplemental materials, the total thickness of the ChG film of $As_{20}Se_{80}$ composition was 50 nm). The comparative analysis of transmission spectra before and after deposition of ChG film shows that the SPR frequency after deposition of ChG film shifted to the red spectral region by 70-150 nm, depending on the thickness of the ChG film and geometric characteristics of the array of gold nanoparticles, and is 680-730 nm (see appropriate curves for pure Au NPs and ChG film/Au NPs structures in Fig. 1 (d)-(e). This fully satisfies the conditions above, since the width of the band gap of ChG films of As-Se system which were researched is $Eg \approx 1.9$-2 eV at the absorption of $\alpha =10^3 см^{-1}$. It corresponds to the laser wavelength of 630-650 nm (shown by arrow in Fig 1 (d)-(e).

**2.3 Experimental Setup** Thus, the layout and composition of the structures are chosen to produce strong light absorption in the visible range resulting in almost complete absorption of incident visible light in the volume. We employed several different designs, but here we will mainly concentrate on one particular structure which showed the best performance (Fig. 2). Two types of experiments were performed.

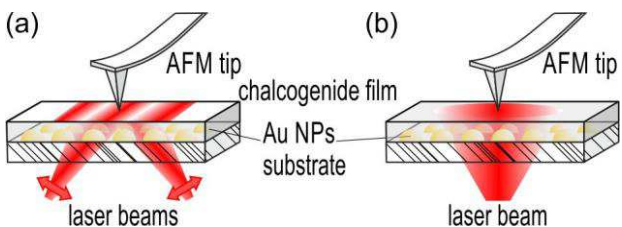

**Figure 2** Scheme of the experimental setup (a) for holographic recording and (b) single-beam exposure. Components along the direction of the laser beam are indicated directly on the images. P-polarized laser source was used for holographic exposure (a) while random polarization (b) was exploited in single-beam experiment.

In the first series of studies (Fig. 2 a) the samples were irradiated from below through a transparent substrate using two coherent beams of solid-state laser with a wavelength of 650 nm (E = 1.9 eV) of equal power density that interfered on the ChG film free surface and produced surface relief grating (SRG). According to [13] the giant vectorial SRG induced by lateral mass transport appears only in the case where the light polarization of the recording beams has a component along the light intensity gradient, so we used p-polarized light (electric fields of both beams are oriented parallel to the plane of interference and parallel to the SRG vector, i.e. perpendicular to the grating lines). In the second series of experiments the samples were irradiated from below through a transparent substrate using unfocused single beam of the same laser of random polarization (Fig 2 b).

Irradiation time ranged from a few minutes to few hours at 80-200 mW/cm$^2$ intensity. This is the level of intensity which is usually used in the formation of surface reliefs in ChG films by of holographic method. Taking into account the peculiarities of *in-situ* measurements of surface deformation by the method of probe microscopy [13], optical recording of the surface relief was studied in real time using AFM (Nanoscope IIIa Dimension 3000, Digital Instruments / Bruker) on a special integrated device which allowed conducting both the light irradiation of the sample and research kinetics of corresponding topographical changes of its surface simultaneously.

**3 Results and discussion**
**3.1 Holographic recording** It is well-known that some Se is extracted from the initial As-Se backbone to react with the photodiffused particles of some noble metals (e.g. Ag) under light irradiation [21]. This is a result of the spontaneous reaction of Ag with some charged metastable states on the chalcogen initiated by light illumination and with charged defects occurring at bond conversion and the two layers being combined by photodissolution. For such a case we could detected both processes (mass-transport due to SPR and photodissolution) simultaneously. To demonstrate the plasmonic nature only onto the results we have check if the photodissolution takes place in case of Au NPs and the series of additional experiments were performed. For this a 300 nm thick $As_{20}Se_{80}$ layer was thermally evaporated in vacuum on the above mentioned Au NPs nanostructures. After that SRG were recorded on the samples by two coherent p-polarized laser beams with equal intensity (Fig. 2 a). Irradiation time was 4 hours.

The electric field of an incoming p-polarized light can induce surface plasmons in the GNP and, possibly, causes their photodissolution in the galssy matrix. As we can see from Fig. 3 (a) the SRG appears. Moreover, SPR triggers a giant mass transport in large enough area (Fig.3, insert) and accelerates SRG formation while increasing it in height up to the order (or even twice more) of the film initial thickness as it follows from AFM image of SRG and its cross-section, see Fig. 3 (b)-(c). Appropriate electron probe microanalysis (EPMA) measurements after the SRG formation was made and no noticeable differences were found in the composition. Only $As_{20}Se_{80}$ compound was detected on the peak surface of SRG while pure gold remained in the valley, so no photodissolution process takes place between chalcogenide matrix and Au NPs.

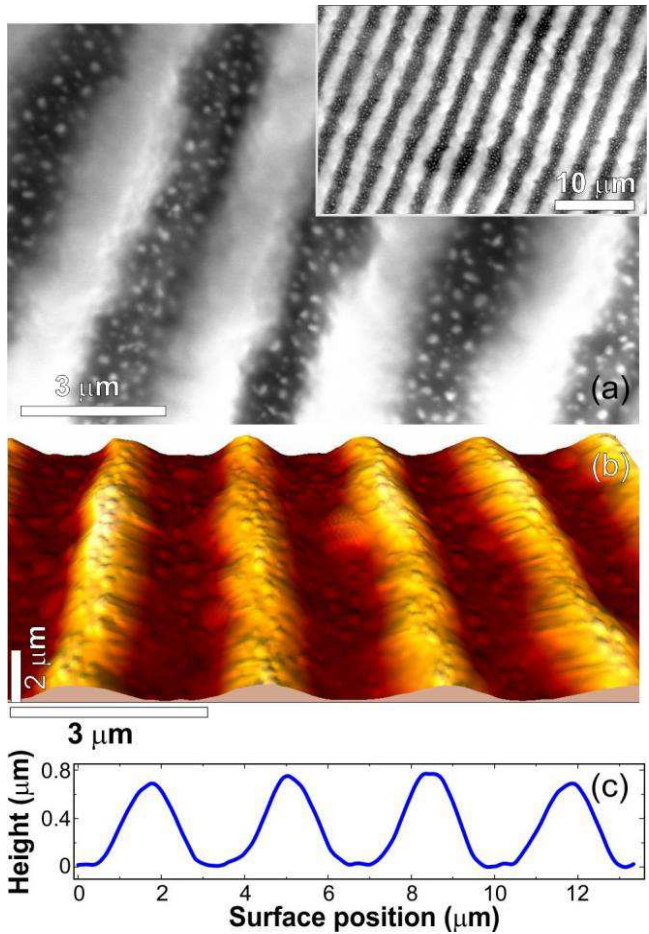

**Figure 3** SEM (a) and AFM (b) images of surface relief grating (3.6 µm periodicity) with appropriate cross-section (c) that was performed in plasmonic structure of 300 nm-thick $As_{20}Se_{80}$ film/Au NPs after 4 hours of recording by 650 nm laser with intensity of 100 mW/cm$^2$. See the text for details.

Note that the same peak-to-valley SRG amplitude (about 700 nm) we can obtain in $As_{20}Se_{80}$ film without plasmonic structures, but under 10 hours of exposure (not shown). It means that enhancement of SRG formation in ChG films through the excitation of plasmonic modes in Au NPs takes place. We note that the behavior of photoinduced SRGs similar to described above has also been observed by us in $As_2S_3$ films [22] also revealing the giant photoplasticity under band-gap illumination [8].

**3.2 Single-beam irradiation** In the AFM research of the reference surface of a sample $As_{20}Se_{80}$/Au NPs before exposure to single-beam light, it was found that the morphology of the array of Au NPs, despite the presence of a thin (100 nm) layer of CG film, can be reliably studied in tapping mode in AFM (see Fig. 4 a and appropriate cross-section in Fig. 4 (d), curve 1). It is clear that such a heterogeneous distribution of Au NPs of different shapes and sizes, which is observed on the reference surface following its irradiation with light, should lead to the excitation of localized surface plasmons with a rather complicated picture of the distribution of the electromagnetic field. That is why we expected no polarization-dependent results and used the laser beam with random polarization. On the other hand, it allows us at this stage of research to simultaneously examine the distribution of the electromagnetic field induced by ensembles of gold nanoparticles of different shapes and sizes.

Fig. 4 (b) shows the topography (AFM image) of the surface of the sample after its irradiation with light for 230 sec. The analysis of dynamics of the changes in the morphology of surface with increasing exposure (Figure 4 b-c, curves 2-3) showed that its roughness (standard deviation - RMS) increased from 1.6 nm for the reference surface to 11.6 nm. Figure 4 (c) also shows that the surface topography has dramatically changed relative to the reference surface and corresponds to the local distribution of the electromagnetic field intensity of surface plasmons due to the concentration of material in the film in the areas of the local maximum of the field. This follows from the general laws of photoinduced mass transport in $As_{20}Se_{80}$ films, including the fact that the movement of material in this composition occurs in the direction towards areas of maximum intensity of light [19]. Thus, according to the obtained cross-sections from the surface, shown in Figure 4 c (curve 3) the height of the protrusions in the areas of maximum intensity of near-field illumination is 14 nm.

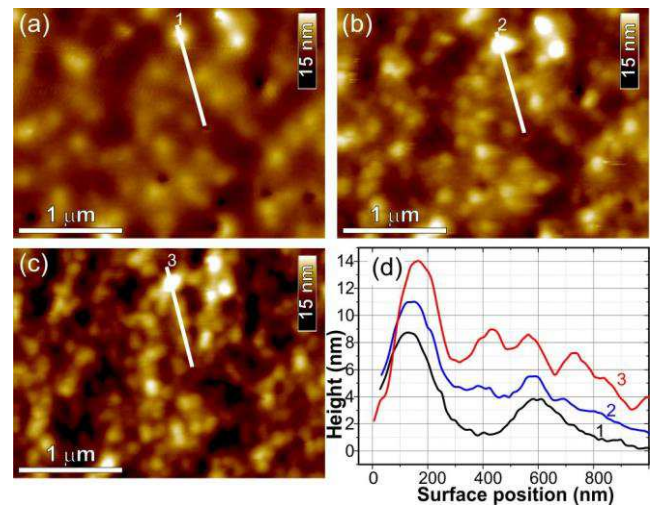

**Figure 4** AFM images of the topography of $As_{20}Se_{80}$ film placed on the Au NPs before irradiation (a). The ChG film thickness is 100 nm. The same area after irradiation with 650 nm laser beam (100 mW/cm$^2$) at 230 s (b) and 625 s (c). Appropriate AFM cross-sections (d) were taken along the lines 1-3 on (a)-(c) and illustrated initial surface topography (curve 1) and surface relief increasing (curves 2-3) due to plasmon intensity distribution into the glass matrix.

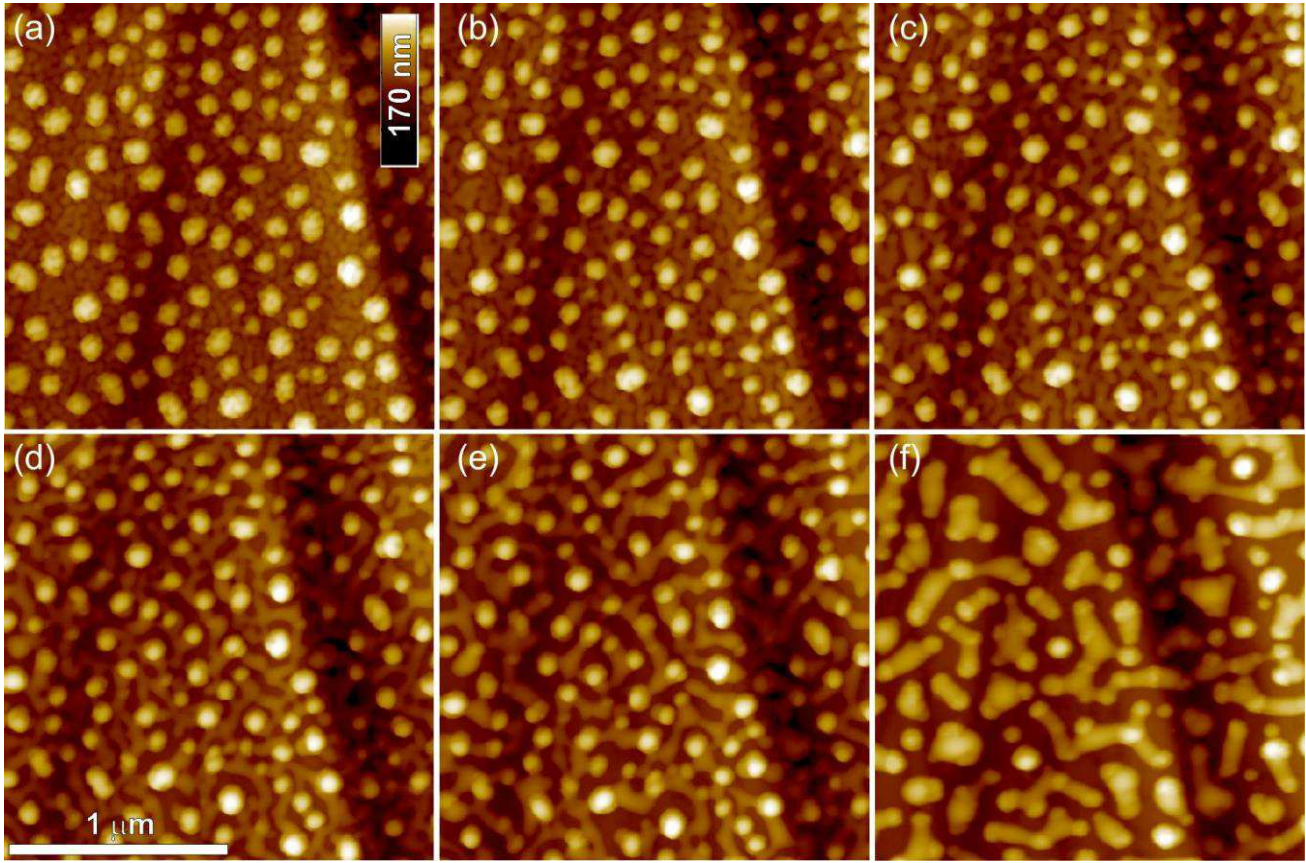

**Figure 5** AFM images of the topography changes of As20Se80 film placed on the Au NPs recorded in-situ during long-term irradiation with 650 nm laser (150 mW/cm2): before irradiation (a); the same area after irradiation at total exposure of 240 s (b), 480 s (c), 1500 s (d), 3000 s (e) and 12 hours (f). The ChG film thickness is 50 nm.

A more detailed analysis of the dynamics of the changes in surface morphology shows that they begin immediately after the laser is turned on (not shown). With increasing exposure the maximum height of the protrusions and surface roughness come to saturation; and after 143 minutes of illumination in the places of local minima of the electromagnetic field of surface plasmons there is observed partial baring of the substrate, which should lead to increase of transparency (photobleaching) of the sample. This fact has been experimentally confirmed in our tests (not shown here).

Similar results were obtained for nano-dimensional (ND) $As_{20}Se_{80}$ film with thickness less than 100 nm deposited on Au NPs with a diameter of 100 nm (Fig 5 a-e). Dramatic change in surface relief was detected under long-term exposure (Fig. 5 f).

At the same time, the opposite result was obtained in the study of a-Se/Au NPs samples (Fig. 6 a). Due to the fact that the direction of photoinduced mass transport in amorphous selenium is opposite to $As_{20}Se_{80}$ composition [20], the material accumulates in the places of the local minimum of light (Figure 6 b-c). This results in a significant reduction in surface roughness of the sample under irradiation with light (from 1.3 nm to 0.6 nm) caused by blurring of the initial picture of the relief (see cross-section in Fig 6 a, curve 1), which was formed by Au NPs on the AFM image (Figure 6 b-c, curves 2-3). It should also be noted that the excitation of localized surface plasmons is always accompanied by a local increase in the ambient temperature [23]. Since amorphous selenium has softening temperature close to room temperature (38 C), the obtained result may be the additive sum of motion of matter due to both photoinduced mass transport to the local minima of light, and through increasing of the fluidity which is the result of local thermal heating.

Note that we omit any numerical simulation and/or fitting of the near-field intensity distributions for comparing it with in situ recorded topography changes (at least in this stage of experiment). This is due to the absence of any ordering in Au NPs array that led to very complex-shape intensity of plasmon radiation with a corresponding complex impact on the photosensitive ChG film and appropriate surface relief. With ordering of Au NPs these relationship awaits further studies and such experiments and a complete model will be discussed with more details in another publication. But, the phenomena described above for ChG of $As_{20}S_{80}$ and a-Se films on Au NPs show the similar dependences upon intensity, spectrum, and exposure time of excitation light as for pure a-Se and $As_{20}S_{80}$ film during mass-transport [13,19,20], which imply the same underlying mechanisms.

The mechanism of the light-induced mass-transport in ChG is still not well studied, despite some attempts to develop a unified model with a complete description of the basic microscopic mechanism. From the macroscopic point of view for phenomenological explanations several models has been proposed [6,24-26]. Among them, the most widely used model is the gradient force model [6] (the model was originally proposed for understanding anisotropic deformations in azobenzene-functionalized polymers [25]) that based on the fact that the electric field gradient of the writing light along the grating vector causes a force on dipoles (dipolar defects or other anisotropic structural units, native or photoinduced, on the scale of about 3 coordination spheres [27] leading to mass-transport due to their interaction and/or rearrangement. Note that the phenomenon occurs far below the glass transition temperature and the thermal processes should be excluded. On the other hand, the light-induced (athermal) softening of glassy matrix takes place (viscosity lowers to $10^{11}$ Pa·s, [12]) that enhances motion of the dipoles under the driving optical force. It means that the gradient moves dipoles (native and/or created by the light) in a matrix softened by the lightening itself. Various spectroscopic studies have shown the existence of short Se-chains in Se-rich glasses [28]. We suggest that short Se segments may act as polarization sensitive anisotropic structural units which can be rearranged under illumination by polarized light in the frame of the mechanism which was adopted to account for photoinduced optical anisotropies [29]. Another possibility for matter motion is photoinduced dipoles, that created by light after scission of the weak bonds of over-coordinated atoms (e.g. hypervalent defects in a-Se, i.e., three-fold and four-fold coordinated Se atoms [30]). Under optical electric field the photoinduced dipoles can lower their energy by changing configuration and/or aligning in the direction along or perpendicular to the polarization of incident light (in case of linear polarization). For both types of dipoles (native and photoinduced) their reorientation, rearrangement and attraction could cause mass-transport only in the presence of driving force and the latter is above mentioned electric-field gradient force. Existence of this force in the case of ChG, however, suffers from difficulties due to some points (see e.g. [31] for details). With this reason, we try to propose other driving force that can causes the mass-transport in ChG. This driving force may arise from anisotropic diffusion of photoexcited carriers leading to appearance of internal electric field. Additional evidence of this hypothesis is delivered by AFM measurements of surface profile and corresponding surface potential that were taken "in situ" under polarized laser irradiation focused in ~ 2 µm spot (Fig.7). Modification of the electrical properties of $As_{20}Se_{80}$ film under irradiation was studied by Kelvin probe force microscopy technique (KPFM), one of electric field sensitive AFM modes, dealing with surface potential [32].

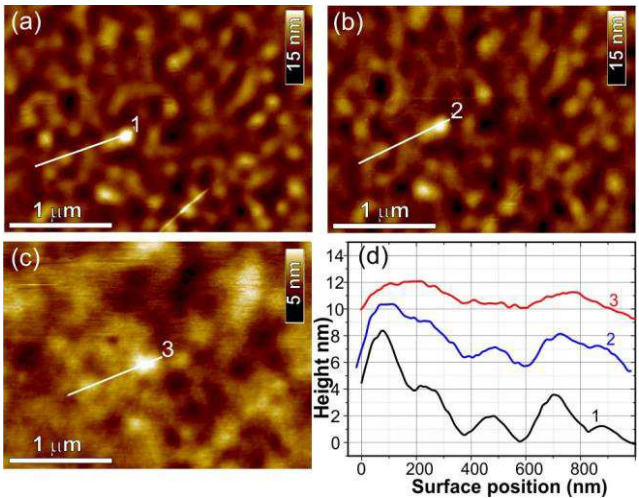

**Figure 6** AFM images of the topography of a-Se film placed on the Au NPs before irradiation (a). The ChG film thickness is 100 nm. The same area after irradiation with 650 nm laser beam (100 mW/cm$^2$) at 29 min (b) and 85 min (c). Appropriate AFM cross-sections (d) were taken along the lines 1-3 on (a)-(c) and illustrated initial surface topography (curve 1) and flattening of the film surface (curves 2-3) due to plasmon intensity distribution into the glass matrix.

Fig 7 a-c show the height of surface deformation versus time of exposure and appropriate KFPM signals (surface potential). Both values are simultaneously appears if the polarized laser irradiation switches on (Fig 7 a) and monotonically increase further with exposure (Fig 7 b-c). Surface potential, however, increases to some saturation level and after that the blurring occurs (Fig 7 e). The maximum of saturated surface potential occurs at 500 s of exposure while the peak-to-valley height h in anisotropic deformations in $As_{20}Se_{80}$ tends to increase with the absorbed dose without explicit saturation (Fig 7 d).

Concerning to the shape of deformation in illuminated spot, one can see that an isotropic photoexpansion appears at first (Fig 7 a), which gradually transforms on the stage of lateral mass-transport to an anisotropic M-shaped deformation with exposure time (Fig 7 b-c). It should be noted that the growth of the anisotropic deformation is delayed with respect to the appropriate M-shaped profile appearing of surface potential. The similar result was obtained for holographic exposure that generates the polarized illumination pattern and the confirmation will be presented elsewhere.

Thus, we conclude that the processes of photomodification of the electric parameters of ChG layers that associated with volume charge appearance and redistribution could be the main reason of the mass-transport. Additionally, we can stimulated further this phenomena with using the local electric field of surface plasmons when ChG cover a metallic (e.g. Au) NPs exposed to light near their SPR plasmon resonance which should to overlap with ChG absorption band.

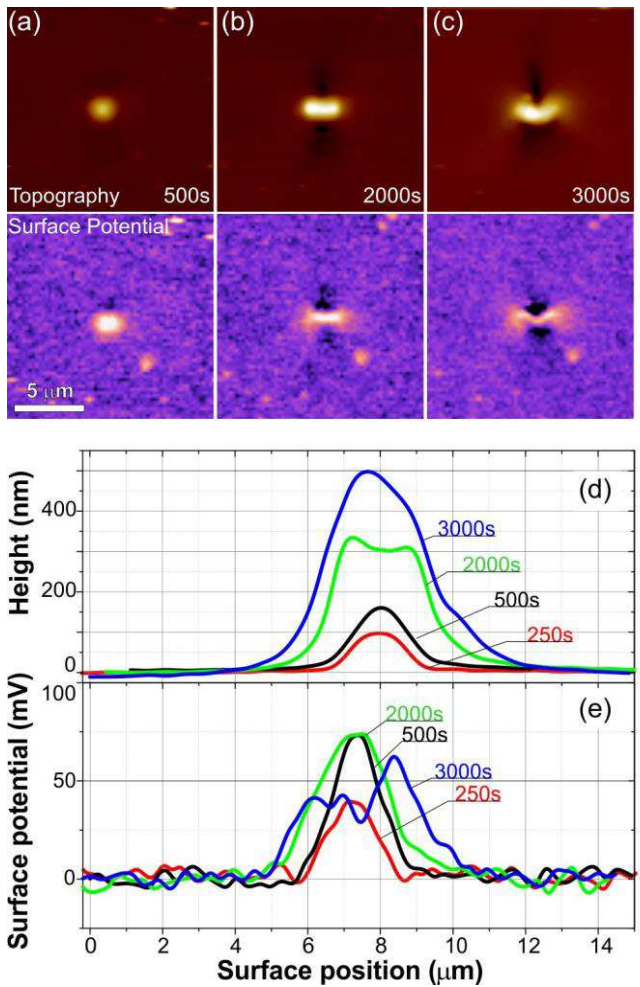

**Figure 7** Surface relief and surface potential distribution as a function of exposure time for irradiated 1μm-thick $As_{20}Se_{80}$ film placed on the glass substrates covered by ITO (a-c); appropriate AFM cross-sections (d-e) oriented perpendicular to the polarization plane. The illumination source is a linearly polarized solid state laser (650 nm), with intensity of ~200 W/cm$^2$ in a focused spot of ~2 μm in diameter. The electric-field direction is vertical.

**4 Conclusion** In this contribution we propose a simple setup for surface plasmon near-field imaging which is based on a photosensitive structure that uses plasmonic signal generation by Au NPs integrated into the non-oxide ChG film matrix or film/substrate interlayer. The influence of near-field illumination on the formation of surface reliefs in films of ChG with integrated Au NPs nanoparticles under the appropriate excitation of SPR by means of laser radiation is investigated.

We have shown that controlled changes in the surface topography of ChG films are possible through near-field illumination which occurs at the excitation of localized surface plasmons. By means of integration of Au NPs in a ChG film, was obtained a corresponding photosensitive structure characterized by an effective overlapping of SPR frequency and the absorption band of ChG. The research of the evolution of surface topography carried out in real time by *in-situ* AFM scanning showed that the topography changes which allows the mapping of surface plasmon intensity distribution. Depending on a ChG composition, the material moves either towards the areas of maximum intensity of light ($As_{20}Se_{80}$), or, respectively, away from them (a-Se). From the obtained results there follows the possibility of changing the surface topography of a ChG film by means of changing the shape, size and geometry of Au NPs. Effects, realised in structures with $As_{20}Se_{80}$ amorphous layer, can be applied for other selected chalcogenide layers and fabrication of locally driven information recording media and mapping of surface plasmon intensity distribution. Some additional possibilities regarding controlled changes of surface topography can be expected through the integration of ordered arrays of Au NPs in ChG that could be intensity and polarization of near-field light.

**Acknowledgements** One of the authors, M.L.T. acknowledges support from International Visegrad Fund. A part of this work was supported by the TAMOP 4.2.2.A-11/1/KONV-2012-0036 project, which is co-financed by the European Union and European Social Fund and by the Ukrainian National Academy of Sciences through the project 6-13H.